\documentclass[aps,eqsecnum,amssymb,floatfix,showpacs,twocolumn]{revtex4}
\usepackage{amsmath,epsfig,bm,pifont}

\usepackage{graphicx} 
\usepackage{bm} 
\usepackage{amssymb}
\usepackage{dcolumn}
\usepackage{amsmath}
\usepackage[normalem]{ulem}
\usepackage[inline]{trackchanges}

\usepackage{perpage}
\MakePerPage{footnote} 

\addeditor{GS}

\begin{document}

\def\gsim{\mathrel{\rlap{\lower4pt\hbox{\hskip1pt$\sim$}}}}
\newcommand{\beq}{\begin{equation}}
\newcommand{\eeq}{\end{equation}}
\newcommand{\beqn}{\begin{eqnarray}}
\newcommand{\eeqn}{\end{eqnarray}}
\newcommand{\btab}{\begin{tabular}}
\newcommand{\etab}{\end{tabular}}
\newcommand{\To}{\Longrightarrow}
\newcommand{\brho}{\mbox{\boldmath$\rho$}}
\newcommand{\bome}{\mbox{\boldmath$\omega$}}
\newcommand{\re}{\nonumber\\}
\newcommand{\etal}{{\em{et al.}}}
\newcommand{\ibid}{{\em{ibid}}}
\newcommand{\tm}{\times}
\newcommand{\lc}{\left<}
\newcommand{\rc}{\right>}
\newcommand{\lr}{\left|}
\newcommand{\rl}{\right|}
\newcommand{\lb}{\left(}
\newcommand{\rb}{\right)}
\newcommand{\ls}{\left[}
\newcommand{\rs}{\right]}
\newcommand{\Lb}{\left\{}
\newcommand{\Rb}{\right\}}
\newcommand{\xchi}{X_\sigma}
\newcommand{\wxchi}{\widetilde{X}_\sigma}
\def\ms{M_s}
\def\mst{\tilde{M}_s}
\newcommand{\qq}{{\bf q}}
\newcommand{\kf}{k_{\rm F}}
\newcommand{\pp}{{\bf p}}
\newcommand{\kk}{{\bf k}}
\newcommand{\hh}{{\bf h}}
\newcommand{\tlq}{\tilde{q}}
\newcommand{\tlt}{\tilde{T}}
\newcommand{\tlo}{\tilde{\omega}}

\newcommand{\tlqq}{\tilde{q}_2}
\newcommand{\tlpp}{\tilde{p}_2}

\newcommand{\gcq}{\, \text{g}/\text{cm}^{3}}

\newcommand{\bea}{\begin{eqnarray}}
\newcommand{\eea}{\end{eqnarray}}

\newcommand{\la}{\langle}
\newcommand{\ra}{\rangle}
\newcommand{\ts}{\textstyle}
\newcommand{\wt}{\widetilde}

{\begin{flushleft} 
LA-UR-12-21688\\
INT-PUB-12-031
\vskip -0.5in
{\normalsize }
\end{flushleft}
\title{Response function of strongly interacting fermi gas in a virial expansion}

\author{Gang Shen} 
\email{gshen@uw.edu}
\affiliation{Theoretical Division, Los Alamos National Laboratory, Los Alamos, NM 87545, USA\\
and Institute for Nuclear Theory, University of Washington, Seattle, WA 98195, USA}

\begin{abstract}
The dynamic response functions of strongly interacting fermion gas in homogeneous space are investigated in a virial expansion to second order. The density response function exhibits transition from atomic to molecular response, as the interaction strength increases and the system undergoes BCS-BEC crossover. 
The qualitative features of density and spin response agree with measurements from the Bragg spectroscopy experiments. 
The virial response is exact at low density and high temperature, therefore providing a benchmark for many-body response.
\end{abstract}
\date{\today}
\maketitle

\section{Introduction}

Recent years the advances in cold atom experiments have improved our understanding of thermodynamic and dynamic properties of strongly interacting systems. Particularly, the dynamic response function gives the response of the system under an excitation probe, therefore describing the inelastic scattering processes. Quite remarkably, the Bragg spectroscopy has been used to measure the dynamic and static density response functions for a strongly interacting cold Fermion system \cite{Veeravalli08}. A smooth transition from atomic to molecular spectra, or from BCS to BEC regimes, is observed with a clear signature of pairing at and above unitarity, which provides a direct link to two-body correlations. Recently, the Bragg spectroscopy is also applied to obtain the dynamic spin response function \cite{spin12}.

Due to its nonperturbative nature, theoretically it is extremely hard to obtain the response function of strongly interacting Fermion system. There have been several investigations on the dynamic response function \cite{DMF,Buchler04,Bruun06,Stringari09,Guo10}. In certain limits, large momentum transfer between external probe and the system \cite{Nishida12}, or system being at very low density, to name a few, the many-body phenomena are expected to be dominated by few-body physics explicitly. The virial expansion presents a tractable approach to the strongly interacting system and has a controllable small parameter, the fugacity $z=\exp(\mu/T)$, when the system is at low density (chemical potential $\mu$) and high temperature ($T$). It has been applied to study the thermodynamic properties of strongly interacting fermi gas \cite{HoMueller,LiuHuDrummond}. For a trapped strongly interacting fermi gas, a quantum virial expansion to second order for the response functions has been developed in Refs. \cite{LiuHuDrummond10b,LiuHu12}.

In this work, the latter virial expansion is extended to study the dynamic response functions of strongly interacting fermi gas in homogeneous space. The formalism closely follows that developed in Ref. \cite{LiuHuDrummond10b} for trapped fermi gas.
The dynamic density response function in homogeneous space is found to exhibit transition from atomic to molecular spectra, as the interaction strength increases and the system undergoes BCS-BEC crossover.  The virial response is exact at low density and high temperature, when fugacity is a small number. Therefore it provides a benchmark for the many-body response functions. %
Qualitatively the response functions of strongly interacting fermi gas in homegeneous space show similar characteristics as those in trapped fermi gas from experiments. They may be related by a local density approximation. In this work, the results in homogeneous space are explicitly written down in compact and closed form with simple integrals. The virial expansion for dynamic response function can be readily applied to other strongly interacting many-body system, like neutron matter and nuclear matter in supernova (see Ref. \cite{HorowitzSchwenk} for a study on the long wavelength behavior of static response function).

The paper is organized as follows. In next section \ref{sec:dynamic} the formalism for the dynamic response function of fermi gas in second order virial expansion is presented in detail. In Sec. \ref{sec:results} the frequency dependence, interaction dependence, and temperature dependence of dynamic response function, as well as the static response function are discussed. The conclusions are given in Sec. \ref{sec:conclusions}.

The natural units $\hbar=k_B=1$ are used throughout.

\section{\label{sec:dynamic}Formalsim for dynamic response functions}

The formalism below closely follows that developed in Ref. \cite{LiuHuDrummond10b} for a trapped fermi gas. Here a recap of the formalism is given for completeness. The dynamic density response for a spin-unpolarized gas under four momentum transfer ($\bf q$,$\omega$) can be separated into two pieces:
\beq\label{sd}
S_D({\bf q},\omega) = 2[S_{\uparrow \uparrow}^{}({\bf q},\omega) + S_{\uparrow \downarrow}^{}({\bf q},\omega)],
\eeq
where $S_{\downarrow \downarrow}^{} = S_{\uparrow \uparrow}^{}$ and 
$S_{\downarrow \uparrow}^{} = S_{\uparrow \downarrow}^{}$ are used. Similarly one could obtain the dynamic spin response function \footnote{This is the correlation function related to spin operator in the quantization direction, $\sigma_z$, as is clear from Eq. (\ref{suscep}).}, 
\beq\label{ss}
S_S({\bf q},\omega) = 2[S_{\uparrow \uparrow}^{}({\bf q},\omega) - S_{\uparrow \downarrow}^{}({\bf q},\omega)].
\eeq
The spin dependent dynamic response functions $S_{\sigma \sigma'}^{}({\bf r}, {\bf r}',\omega)$ in above two equations (a simple Fourier transform with respect to ${\bf r}- {\bf r}'$ then gives $S_{\sigma \sigma'}^{}({\bf q},\omega)$) can be obtained via the fluctuation-dissipation theorem and analytic continuation from the dynamic susceptibility,
\beq\label{fd}
S_{\sigma \sigma'}^{}({\bf r}, {\bf r}',\omega) = 
\frac{-\text{Im}\chi^{}_{\sigma \sigma'}({\bf r}, {\bf r}'; i \omega_n \to \omega + i0^{+})}
{\pi(1 - e^{-\beta \omega})},
\eeq
where the matsubara frequencies $\omega_n = 2\pi n k_BT$ ($n=0,\pm1,...$), and the time-dependent correlation function in imaginary time is defined as usual, 
\beq\label{suscep}
\chi_{\sigma \sigma'}({\bf r}, {\bf r}', \tau) = 
-\langle 
T_\tau^{} 
\hat n_{\sigma}({\bf r},\tau)
\hat n_{\sigma'}({\bf r},0)
\rangle,
\eeq
where $T_\tau^{}$ is the imaginary-time ordering operator, and $\hat n_{\sigma}({\bf r},\tau)$ is the density (fluctuation) operator in spin channel $\sigma$, and $\tau$ is an imaginary time in the interval $0<\tau\le\beta=1/k_BT$.
By the usual virial expansion in fugacity $z=e^{\beta\mu}$,
\bea
\chi^{}_{\sigma \sigma'} 
&=& z X^{}_1[1 + z (X^{}_2/X^{}_1  - Q_1) + \dots],
\eea
where 
\bea
Q_N\ &=& \text{Tr}_N \left[ e^{-\beta \hat H} \right], \\
X^{}_N &=& 
-\text{Tr}^{}_N
\left[
e^{-\beta \hat H}
e^{\tau \hat H}
\hat n_{\sigma}({\bf r})
e^{-\tau \hat H}
\hat n_{\sigma'}({\bf r'})
\right].
\eea
$\hat H$ is the Hamiltonian. The lower script $N$ indicates the trace is taken over $N$-body state.
The interaction enters from second order term. Using completeness relations, 
\bea
&&\Delta \chi^{}_{\sigma \sigma',2}({\bf r},{\bf r'},\tau) = \Delta X^{}_2 = X^{}_2 - X^{0}_2 \cr
&&=
- \sum_{P,Q} 
\left[
e^{-\beta E^{}_P + \tau (E^{}_P - E^{}_Q)}
C^{PQ}_{\sigma \sigma'}({\bf r},{\bf r}')
\right], 
\eea
where the superscript $``0"$ in $X_2^0$ indicates quantities for non-interacting system, and 
\beq
C^{PQ}_{\sigma \sigma'}({\bf r},{\bf r}') =
\langle P |\hat n_{\sigma}({\bf r})|Q \rangle
\langle Q | \hat n_{\sigma'}({\bf r}')| P \rangle,\
\eeq
pair state $(\uparrow\downarrow)$ $| P \rangle$ ($| Q \rangle$) has energy $E^{}_P$ ($E^{}_Q$).
One can perform discrete Fourier transform on $\Delta \chi_{\sigma \sigma',2}({\bf r},{\bf r'},\tau) $,
and analytically continuate the results via $i\omega_n \rightarrow \omega+i\eta$,
to get the second order response function,
\bea
&&\Delta S_{\sigma \sigma',2}({\bf r},{\bf r'},\omega)\cr 
&=& -\frac{\text{Im}\Delta \chi_{\sigma \sigma',2}({\bf r},{\bf r'},\omega)}{(1-e^{-\beta\omega})\pi} \cr
&=& \sum\limits_{P,Q} \delta(\omega+E_P-E_Q)e^{-\beta E_P} C_{\sigma\sigma'}^{PQ}({\bf r},{\bf r'}).
\eea
Through a further Fourier transform the momentum space response function follows as,
\bea
&&\Delta S_{\sigma \sigma',2}({\bf q},\omega)\cr
&=& \sum\limits_{PQ}\delta(\omega+E_P-E_Q) e^{-\beta E_P} F_{\sigma\sigma'}^{PQ}({\bf q}),
\eea
where, 
\bea
F_{\sigma\sigma'}^{PQ}({\bf q})\ =\ \int d{\bf r}d{\bf r'}e^{-i{\bf q}({\bf r}-{\bf r'})} C_{\sigma\sigma'}^{PQ}({\bf r},{\bf r'}).
\eea
After some algebra, one can arrive at following form for the response function,
\bea\label{s2}
\Delta S_{\sigma\sigma^{\prime},2}({\bf q},\omega)=\int d\omega^{\prime}W_{cm}({\bf q},\omega^{\prime})W_{rel}^{\sigma\sigma^{\prime}}({\bf q},\omega-\omega^{\prime}),
\eea
with following notations for the center of mass piece $W_{cm}({\bf q},\omega^{\prime})$ and relative motion piece $W_{rel}^{\sigma\sigma^{\prime}}({\bf q},\omega-\omega^{\prime})$ contributions, respectively.
\bea 
W_{cm}({\bf q},\omega^{\prime})\equiv\sum_{p1q1}\delta(\omega^{\prime}+\epsilon_{p1}-\epsilon_{q1})e^{-\beta\epsilon_{p1}}\left|f_{p1q1}\right|^{2},
\eea and,
\bea
f_{p1q1}\equiv\int d{\bf R}e^{-i{\bf q}\cdot{\bf R}}\varphi_{p1}^{*}({\bf R)}\varphi_{q1}({\bf R)},
\eea
where $\varphi_{{\bf p1},{\bf q1}}({\bf R})$ is plane wave state for center of mass motion of two-fermion with energy of $\epsilon_{p1}, \epsilon_{q1}$. 
\bea\label{wrel}
&&W_{rel}^{\sigma\sigma^{\prime}}({\bf q},\omega-\omega^{\prime})\cr &&\equiv\sum_{p2q2}
\delta(\omega-\omega^{\prime}+\epsilon_{p2}-
\epsilon_{q2})e^{-\beta\epsilon_{p2}}A_{p2q2}^{\sigma\sigma^{\prime}},
\eea
with 
$A_{p2q2}^{\uparrow\uparrow}\equiv\left|A_{p2q2}\right|^{2}$ and
$A_{p2q2}^{\uparrow\downarrow}\equiv A_{p2q2} A_{q2p2}$, and 
\beq
A_{p2q2}\equiv\int d{\bf x}e^{-i{\bf q}\cdot{\bf x}/2}\phi_{p2}^{*}({\bf x)}\phi_{q2}({\bf x)},
\eeq
where $\phi_{{\bf p2},{\bf q2}}({\bf x})$ is state for relative motion of two-fermion with energy of $\epsilon_{p2}, \epsilon_{q2}$. 

We mention by passing the center of mass piece 
\beq\label{wcm}
W_{cm}({\bf q}, \omega')\equiv\frac{m^2}{n\pi^2\beta q} e^{-(\omega'-q^2/4m)^2\beta m/q^2},
\eeq where $n$ is the number density and $m$ is the atomic mass.
In the next the resulting second order response functions are presented from BCS (\ref{bcs}) to BEC (\ref{bec}) side.

\subsection{\label{bcs}BCS side}

Using the $S$-wave scattering state wave function $\phi_S({pr})\ =\ \sqrt{2/\pi(1+p^2a^2)}\left[\sin (pr)-pa\cos (pr)/pr\right]Y_{00}$ \footnote{There should be a factor of $e^{-i\delta_0}$ where $\delta_0$ is the phase shift. However this overall phase does not influence the results in the paper.} ($a$ is the $S$ wave scattering length), the relative piece in Eq. (\ref{wrel}) can be written as,
\begin{widetext}
\bea
W_{rel}^{\sigma\sigma^{\prime}}(q,\omega) &=& \int dp_2 p_2^2 dq_2 q_2^2 \delta({\omega+\epsilon_{p_2}-\epsilon_{q_2}})e^{-\beta\epsilon_{p_2}} \biggl[ \left(\int dx x^2 j_0(qx/2) \phi_S(p_2x) \phi_S(q_2x) \right)^2 \cr 
&&
+ \sum\limits_{l>0} (2l+1) (-1)^{l(1-\delta_{\sigma\sigma'})} \left( \int dx x^2 j_l(qx/2)\sqrt{\frac{2}{\pi}} j_l(p_2x)\phi_S(q_2x) \right)^2 \cr
&&
+ \sum\limits_{l>0} (2l+1) (-1)^{l(1-\delta_{\sigma\sigma'})} \left( \int dx x^2 j_l(qx/2)\sqrt{\frac{2}{\pi}} j_l(q_2x)\phi_S(p_2x) \right)^2 
-\mathrm{ non.\ inter.\ terms} \biggr].
\eea 
\end{widetext}
Note either $\phi_S(p_2x)$ or $\phi_S(q_2x)$ has to be interacting $S$ wave state, otherwise it will cancel with corresponding non-interacting term.
With above equation and Eq. (\ref{wcm}) one can proceed to the dynamic structure function in Eq. (\ref{s2}),

\begin{widetext}
\bea\label{bcs1}
\Delta \tilde{S}_{\sigma\sigma^{\prime},2}(q,\omega) &=& \int d\tlpp \tlpp^2 d\tlqq \tlqq^2 \frac{3\tlt}{4\tlq} e^{-(\tlo+2\tlpp^2-2\tlqq^2-\tlq^2/2)^2/2\tlq^2\tlt-2\tlpp^2/\tlt} \biggl[ \left(\int dx x^2 j_0(\tlq x/2) \phi_S(\tlpp x) \phi_S(\tlqq x) \right)^2 \cr 
&&
+ \sum\limits_{l>0} (2l+1) (-1)^{l(1-\delta_{\sigma\sigma'})} \left( \int dx x^2 j_l(\tlq x/2)\sqrt{\frac{2}{\pi}} j_l(\tlpp x)\phi_S(\tlqq x) \right)^2 \cr
&&
+ \sum\limits_{l>0} (2l+1) (-1)^{l(1-\delta_{\sigma\sigma'})} \left( \int dx x^2 j_l(\tlq x/2)\sqrt{\frac{2}{\pi}} j_l(\tlqq x)\phi_S(\tlpp x) \right)^2 
- \mathrm{non.\ inter.\ terms} \biggr],
\eea \end{widetext}
where scaled variables, momentum in terms of $k_F$ and energy in terms of $\epsilon_F=k_F^2/2m$, have been used. For instance, $\tlq=q/k_F$, $\tlt=T/\epsilon_F$, and $\Delta \tilde{S}_{\sigma\sigma^{\prime},2}=\Delta S_{\sigma\sigma^{\prime},2}\times\epsilon_F$. Note the subtraction of non-interacting piece is essential to obtain physical finite results (See App. \ref{app:app2} for more discussions).

After some straightforward algebra, one could obtain the dynamic response functions on the BCS side as follows,
\begin{widetext}
\bea\label{sbcs}
 \tilde{S}^{BCS}_{\uparrow\downarrow}(q,\omega)\ &=&\  z^2(\Delta\tilde{S}^a_{\uparrow\downarrow,2}(q,\omega)+ \Delta\tilde{S}^b_{\uparrow\downarrow,2}(q,\omega)),  \\
 \label{sbcs2} \tilde{S}^{BCS}_{\uparrow\uparrow}(q,\omega)\ &=&\ \tilde{S}_{F}(q,\omega)\ +\  z^2(\Delta\tilde{S}^a_{\uparrow\uparrow,2}(q,\omega)+ \Delta\tilde{S}^b_{\uparrow\uparrow,2}(q,\omega)), 
\eea
\end{widetext}
where terms with superscript ``$a$" indicate contribution from scattering between $S$ wave initial and final states, and ``$b$" for scattering between $l > 0$ wave and $S$ wave states. For details see Eqs. (\ref{bcs_s}), (\ref{bcs_ud}), and (\ref{bcs_uu}) in App. \ref{app:m}.
$\tilde{S}_{F}(q,\omega)$ is the dynamic response function for non-interacting fermi gas \cite{Mazzanti},
\beq\label{sfree}
\tilde{S}_{F}(q,\omega)\ =\ \frac{3\tlt}{16\tlq}\frac{1}{1-e^{-\tlo/\tlt}}\log\left[ \frac{1+ze^{-(\tlo/\tlq-\tlq)^2/4\tlt}}{1+ze^{-(\tlo/\tlq+\tlq)^2/4\tlt}} \right].
\eeq
Note a factor of half is included in above equation to remove spin degeneracy.

\subsection{\label{bec}BEC side}

On BEC side there exist contributions from bound state $\phi_b({\bf r}) =e^{-r/a}/r\sqrt{2\pi a}$ besides scattering states, where binding energy $E_b= -1/ma^2$. An important contribution on BEC side comes from transition between bound states, or molecular response,
\begin{widetext}
\bea\label{bec2}
 \Delta  \tilde{S}^c_{\sigma\sigma^{\prime},2}(q,\omega) &=& \frac{3\tlt}{4\tlq} e^{-(\tlo -\tlq^2/2)^2/2\tlq^2\tlt+2/\tlt k_F^2a^2}\left[ \int dx x^2 j_0(\tlq x/2) 4\pi \phi_b(x)^2\right]^2 \cr
&=&  \frac{3\pi\tlt}{2\tlq^2 k_Fa} e^{-(\tlo -\tlq^2/2)^2/2\tlq^2\tlt+2/\tlt k_F^2a^2} (1+\tlq^2 k_F^2a^2)^{-1/2} P_{-1/2}^{-1/2}[(1+\tlq^2 k_F^2a^2)^{-1/2} ]^2.
\eea
\end{widetext}
It is clear from Eq. (\ref{bec2}) that on deep BEC side, $k_Fa \ll 1 $, the molecular response dominates in the dynamic response function.

Summing up Eqs. (\ref{bcs_s}) with (\ref{bcs_ud}) or (\ref{bcs_uu}), as well as (\ref{bec1}) and (\ref{bec2}), one could obtain the dynamic response functions on the BEC side as follows,
\begin{widetext}
\bea\label{sbec}
 \tilde{S}^{BEC}_{\uparrow\downarrow}(q,\omega)\ &=&\  z^2(\Delta\tilde{S}^a_{\uparrow\downarrow,2}(q,\omega)+ \Delta\tilde{S}^b_{\uparrow\downarrow,2}(q,\omega)+\Delta\tilde{S}^c_{\uparrow\downarrow,2}(q,\omega)+\Delta\tilde{S}^d_{\uparrow\downarrow,2}(q,\omega)),  \\
\label{sbec2} \tilde{S}^{BEC}_{\uparrow\uparrow}(q,\omega)\ &=&\ \tilde{S}_{F}(q,\omega)\ +\  z^2(\Delta\tilde{S}^a_{\uparrow\uparrow,2}(q,\omega)+ \Delta\tilde{S}^b_{\uparrow\uparrow,2}(q,\omega)+\Delta\tilde{S}^c_{\uparrow\uparrow,2}(q,\omega)+\Delta\tilde{S}^d_{\uparrow\uparrow,2}(q,\omega)), 
\eea
\end{widetext}
where the terms with superscript ``$d$" indicate the contribution from transition between bound state and scattering state, shown in Eq. (\ref{bec1}) of App. \ref{app:m}.


\section{\label{sec:results}Results and Discussions}

In this section, the dependences of dynamic response function on frequency, interaction, and temperature, as well as static response function are discussed in details.
The Bragg spectroscopy experiments have been carried out using large momentum transfer. Therefore in most of this section,  the response functions are calculated with a large momentum transfer $q=3k_F$.

\subsection{Frequency dependence of ${S}_{\uparrow\uparrow}$ and ${S}_{\uparrow\downarrow}$}

In Figure \ref{fig:1},  the dynamic response functions for spin-parallel and spin-anti-parallel cases, $S_{\uparrow\uparrow}$ and $S_{\uparrow\downarrow}$, are shown as function of frequency at different interaction strengths for $T=2T_F$ (left panels) and $T=3T_F$ (right panels). In all the results, the largest fugazity is $z$=0.25 (BCS side $1/k_Fa=-1$ and $T/T_F$=2). These small values suggest that higher order corrections to the virial expansion will be small. $\omega_R=q^2/2m$ is the atomic recoil frequency. 

The effect of interaction is most clear in the dynamic response function for spin-anti-parallel case,  ${S}_{\uparrow\downarrow}$, as Eqs. (\ref{sbcs}, \ref{sbec}) contain only interaction induced contributions. On the BCS side $1/k_Fa=-1$, ${S}_{\uparrow\downarrow}$ is peaked around the molecular recoil frequency $\omega_R/2$ due to strong pair correlation from attractive interaction. The peak becomes more visible as temperature decreases and/or interactions strength increases. One notes that there is no peak around atomic recoil frequency $\omega_R$, since the zero-range interaction makes two spin-anti-parallel atoms tightly correlated even at very high momentum transfer, giving rise to a molecular response peak. This is clearly in agreement both with experimental finding for the same response function of fermi gas at low temperature in Ref.~\cite{spin12}, and with theoretical calculations in Ref.~\cite{LiuHuDrummond10b} for trapped fermi gas at high temperature. Furthermore, there is a semi-analytical way to interpret the molecular peak in the spin-anti-parallel response, by using a random phase approximation (RPA) type calculation as follows, 
\beq
S^{\rm{RPA}}_{\uparrow\downarrow}(q,\omega)\ =\ \frac{1}{1-{\rm{exp}}^{-\omega/T}} \rm{Im} \biggl[ \frac{\Pi_0}{1-V_0\Pi_0} - \frac{\Pi_0}{1+V_0\Pi_0} \biggr],
\eeq
where $V_0$ is the $S$-wave interaction between spin up and spin down fermions (for example some modified P\"{o}schl-Teller potential \cite{Forbes12}), and $\Pi_0$ is the Lindhard function for the polarization function of free fermi gas that could be obtained from Eq.~(\ref{sfree}). In the limit that finite range effect in $V_0$ is negligible when momentum transfer is not large enough, one could show the RPA response function $S^{\rm{RPA}}_{\uparrow\downarrow}$ peaks around the molecular response, because the external probe can only excite the short-range strongly correlated pair together.

On the BEC side, the spin-anti-parallel response is sharply peaked around the molecular recoil frequency due to the formation of bound pair state, which dominates the response function.
The response function ${S}_{\uparrow\uparrow}$, given in Eqs. (\ref{sbcs2}, \ref{sbec2}), receives contribution from non-interaction terms Eq.~(\ref{sfree}), which peaks at the atomic recoil frequency. Note the contribution from interaction also peaks around molecular response as the spin-anti-parallel response above. On the BCS side, the interaction is too weak so the response is dominated by the non-interacting terms. On the BEC side, the peak of response function ${S}_{\uparrow\uparrow}$ moves toward the molecular recoil frequency as  ${S}_{\uparrow\downarrow}$ because of formation of bound pair state.

\begin{figure}[th]
\includegraphics[width=1.05\columnwidth]{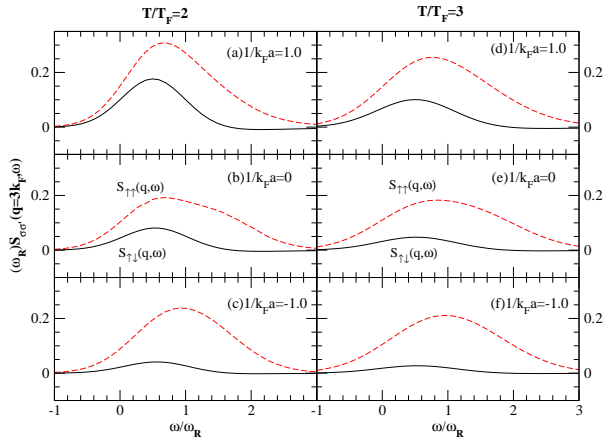}
\caption{(Color online) $S_{\uparrow\uparrow}(q=3k_F,\omega)$ and $S_{\uparrow\downarrow}(q=3k_F,\omega)$ as function of frequency at different $1/k_Fa$ and at $T=2T_F$ (left) and $T=3T_F$ (right).}
 \label{fig:1}
\end{figure}

\subsection{Interaction dependence of dynamic response function}

In Figure \ref{fig:3} the dynamic density (left) and spin (right) responses at $T=3T_F$, obtained by Eqs. (\ref{sd}, \ref{ss}),  are shown for various interaction strengths. The spin response has broad peak around atomic frequency both on BEC and BCS sides, which agrees with the experiment \cite{spin12}. The density response function peaks at atomic recoil frequency on BCS side. As interaction increases, the peak becomes red-shifted. On BEC side, the density response becomes peaked around molecular frequency. These features qualitatively agree with recent Bragg spectroscopy experiment \cite{spin12}, although the latter experiment was performed at extremely low temperature in the superfluid phase. Note, however, here at $T=3T_F$ the density response at unitarity has only a single peak in contrast to double peak found in experiments \cite{spin12} for system in superfluid phase. 

\begin{figure}[th]
\includegraphics[width=1.05\columnwidth]{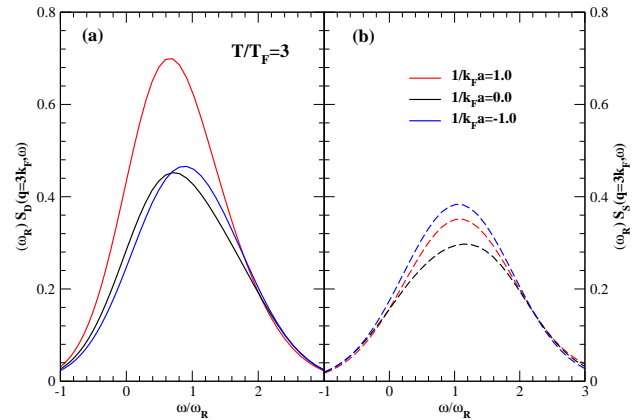}
\caption{(Color online) Interaction dependence of dynamic density (left) and spin (right) responses at $T=3T_F$. }
 \label{fig:3}
\end{figure}

Virial expansion may not be applicable at same temperatures as in the experiments \cite{Veeravalli08,spin12} performed below the critical temperature in the superfluid phase, but around $T_F$ and at unitarity the virial density response already exhibits double peak features similar to experimental findings. In left panel of Fig.~\ref{fig:6}, the density response function at unitarity is shown as function of $\omega/\omega_R$ at temperatures $T=T_F, 1.25 T_F, 1.5 T_F$, and $q=3k_F$. As temperature decreases, the density response function changes from single peak at molecular response (0.5$\omega_R$) for $T=1.5T_F$, to develop an additional shallow peak at $\sim 1.5\omega_R$ for $T=T_F$. The second peak moves to lower frequency as the momentum transfer increases to $q=4.5k_F$ (value used in the Bragg spectroscopy experiment \cite{spin12}), as shown in right panel of Fig.~\ref{fig:6} where $T=T_F$ and at unitarity. At $T=T_F$ and unitarity the fugacity $z$=0.49, so one should view these results with caution and expect sizable correction from higher order terms. Nevertheless, the qualitative feature of density response with double peak at unitarity and $T=T_F$ is similar to what was observed in the experiment \cite{spin12}. It is possible that with higher order virial expansion the second peak would be closer to the atomic response $\omega_R$, and this warrants further studies. Finally one notes that in a previous work using virial expansion for trapped fermi gas \cite{LiuHuDrummond10b}, the density response function only has a single peak even at a temperature of $0.5T_F$, in contrast to experimental finding \cite{spin12} and current work. It is possible that, at temperature lower than $0.5T_F$ the virial expansion for trapped system may also give rise to double peak feature for the density response function.

\begin{figure}[th]
\includegraphics[width=1.05\columnwidth]{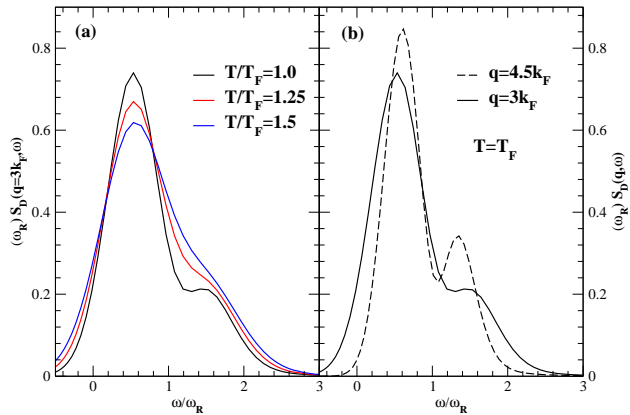}
\caption{(Color online) Left panel: at unitarity the density response function as function of $\omega/\omega_R$ at lower temperatures $T=T_F, 1.25 T_F, 1.5 T_F$ and $q=3k_F$. Right panel: at unitarity the density response function as function of $\omega/\omega_R$ at $T=T_F$, and with different $q=3k_F, 4.5k_F$.}
\label{fig:6}
\end{figure}

\subsection{Static response function}

\begin{figure}[th]
\includegraphics[width=1.05\columnwidth]{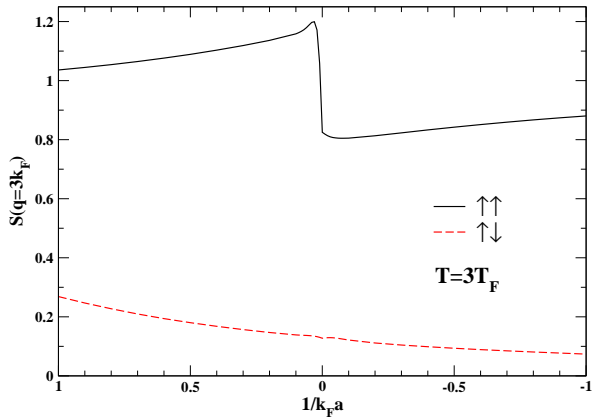}
\caption{(Color online) Static response functions ${S}_{\uparrow\uparrow}$ (q=3$k_F$) and ${S}_{\uparrow\downarrow}$ (q=3$k_F$) as function of $1/k_Fa$ at $T=3T_F$. }
\label{fig:5}
\end{figure}

The static response functions are obtained via following integrals, 
\beq\label{static}
S_{\sigma\sigma'}(q)\ =\ 2\int_{-\infty}^{\infty} d\omega S_{\sigma\sigma'} (q,\omega).
\eeq 
In Figure \ref{fig:5} the static response functions ${S}_{\uparrow\uparrow}$ (q=3$k_F$) and ${S}_{\uparrow\downarrow}$ (q=3$k_F$) are shown as function of $1/k_Fa$ at $T=3T_F$. ${S}_{\uparrow\uparrow}$ (q=3$k_F$) increases from about one in deep BEC side when $1/k_Fa$ reduces, and it decreases quickly approaching unitarity. The rise and fall is due to contribution from transition between bound state and scattering state on the BEC side in Eq. (\ref{bec1}), which vanishes at unitarity. On the other hand,  ${S}_{\uparrow\downarrow}$ (q=3$k_F$) changes more smoothly from BEC to BCS side. Note at unitarity when $T=3T_F$, the fugacity is about 0.13, therefore one expects the higher order terms contribute at about 10-20\% of second order term displayed here.

\section{\label{sec:conclusions}Conclusions}

In this work, the dynamic response functions of strongly interacting fermi gas in homogeneous space are investigated in a virial expansion to second order. The dynamic density response function is found to exhibit transition from atomic response to molecular response, as the interaction strength increases and the system undergoes BCS-BEC crossover. The spin response function has a broad peak around atomic recoil frequency at high temperature. These features qualitatively agree with recent experiments via Bragg spectroscopy. The virial response is exact at low density and high temperature, when fugacity is a small number.  The fugacity discussed in this work is relatively small. Therefore it provides a benchmark for the many-body response functions.

Qualitatively, the response functions of strongly interacting fermi gas in homegeneous space show similar characteristics as those in trapped fermi gas. In fact they may be related by a local density approximation, which would be interesting to study in future work.

To second order virial expansion, the dynamic response function in homogeneous space can be explicitly written down in compact and closed form with simple integrals. This illustrates that the virial expansion to response function may be also applied to other strongly interacting many-body system at low density and high temperature, like neutron matter and nuclear matter in supernova.

It is a pleasure to acknowledge the encouragement from Joe Carlson, Chuck Horowitz, and Sanjay Reddy, as well as helpful interactions with G. Bertsch, C.-C. Chien, J. Drut, M. Forbes, S. Gandolfi, and Y. Nishida. 
The work is supported by a grant from the DOE under contract DE-AC52-06NA25396 and the DOE topical collaboration to study ``Neutrinos and nucleosynthesis in hot and dense matter".

\appendix

\section{Sum of product of two Legendre Polynomials}
\label{app:app2}

The infinite sum over product of two Legendre polynomials  can be carried out utilizing following relations,
\bea
&&\sum\limits_{l\ge 0} (-1)^l (2l+1) P_l(x)^2=\delta(x),\ \ x\in [-1,1], \\
&&\sum\limits_{l\ge 0}  (2l+1) P_l(x) Q_l(x) = 0 , \\
&&\sum\limits_{l\ge 0} (-1)^l (2l+1) P_l(x) Q_l(x) = \frac{1}{2x} , x\in [-1,1], \\
&&\sum\limits_{l\ge 0} (-1)^l (2l+1)Q_l(x)^2 = \frac {1} {2x} \log|\frac{1+x}{1-x}|, \\
&&\label{q2p2} \sum\limits_{l\ge 0} (2l+1) Q_l(x)^2 = (\frac{\pi}{2})^2\sum\limits_{l\ge 0} (2l+1) P_l(x)^2+\frac{1}{x^2-1}. \cr
&&
\eea
Note in Eq. (\ref{q2p2}) the two sums are separately divergent when $x\in(-1,1)$, but their difference is finite. The physical origin of the divergence comes from that the non-interacting two-body piece is proportional to volume times a single particle response function. Therefore it is important to subtract the non-interacting piece at the same (second) order, which serves as a counter term.

\section{Terms (a), (b), (d) for second order response function}
\label{app:m}

The first term in Eq. (\ref{bcs1}) comes from scattering between $S$ wave initial and final states and can be worked out as follows,
\begin{widetext}
\bea\label{bcs_s}
\Delta  \tilde{S}^a_{\sigma\sigma^{\prime},2}(q,\omega) &=& \int d\tlpp  d\tlqq  \frac{3\tlt}{4\tlq^3} (\frac{2}{\pi})^2 e^{-(\tlo+2\tlpp^2-2\tlqq^2-\tlq^2/2)^2/2\tlq^2\tlt-2\tlpp^2/\tlt} \times \cr
&&  \biggl[ \frac{\pi}{4}C_{p2}C_{q2} \bigl(-\mathrm{sign}(\tlq/2-\tlpp-\tlqq)+\mathrm{sign}(\tlq/2+\tlpp-\tlqq)+\mathrm{sign}(\tlq/2-\tlpp+\tlqq)-\mathrm{sign}(\tlq/2+\tlpp+\tlqq) \bigr)  \cr
&&  + \frac{\pi}{4}S_{p2}S_{q2} \bigl(\mathrm{sign}(\tlq/2-\tlpp-\tlqq)+\mathrm{sign}(\tlq/2+\tlpp-\tlqq)+\mathrm{sign}(\tlq/2-\tlpp+\tlqq)+\mathrm{sign}(\tlq/2+\tlpp+\tlqq) \bigr) \cr
&& +\ \frac 1 4 C_{p2}S_{q2} \log\frac{(1+v_2)^2}{(1-v_2)^2} +\ \frac 1 4 C_{q2}S_{p2} \log\frac{(1+v_3)^2}{(1-v_3)^2}
\biggr]^2 - \mathrm{non.\ inter.\ terms},
\eea
\end{widetext}
where sign() is the sign function, and 
$C_{p2}=1/\sqrt{1+\tlpp^2 k_F^2a^2}, 
S_{p2}=-\tlpp k_Fa/\sqrt{1+\tlpp^2 k_F^2a^2}, 
v_2=\left[\tlq^2/4+\tlpp^2-\tlqq^2\right]/\tlq\tlpp, 
v_3=\left[\tlq^2/4+\tlqq^2-\tlpp^2\right]/\tlq\tlqq. $

The second and third terms in Eq. (\ref{bcs1}) come from transition between $l > 0$ partial waves and $S$ wave, and are simplified as follows,
\begin{widetext}
\bea\label{bcs2}
\Delta  \tilde{S}^b_{\sigma\sigma^{\prime},2}(q,\omega) &=&\ \int d\tlpp  d\tlqq  \frac{3\tlt}{4\tlq^3} (\frac{2}{\pi})^2 e^{-(\tlo+2\tlpp^2-2\tlqq^2-\tlq^2/2)^2/2\tlq^2\tlt-2\tlpp^2/\tlt} \sum\limits_{l>0} (2l+1) (-1)^{l(1-\delta_{\sigma\sigma'})}  \times\cr
&& \biggl[ (\frac{\pi}{2})^2P_l(v_2)^2(C_{q2}^2-1)+Q_l(v_2)^2S_{q2}^2+\pi P_l(v_2)Q_l(v_2)C_{q2}S_{q2} \cr
&& +\ (\frac{\pi}{2})^2P_l(v_3)^2(C_{p2}^2-1)+Q_l(v_3)^2S_{p2}^2+\pi P_l(v_3)Q_l(v_3)C_{p2}S_{p2}
\biggr],
\eea
\end{widetext}
where $P_l$ and $Q_l$ are $l^{th}$-order Legendre polynomials of first and second kind, respectively. Note except $(Q_l)^2$ terms, the other terms are nonzero only for $|v_2|$ or $ |v_3| \le 1$ \cite{foot1}. Using the summation relation for the product of two Legendre polynomials (see Appendix \ref{app:app2}), these terms can be further simplified as follows.
\begin{enumerate}
\item Spin anti-parallel case
\begin{widetext}
\bea\label{bcs_ud}
\Delta  \tilde{S}^b_{\uparrow\downarrow,2}(q,\omega) &&=\ \int d\tlpp  d\tlqq  \frac{3\tlt}{4\tlq^3} (\frac{2}{\pi})^2 e^{-(\tlo+2\tlpp^2-2\tlqq^2-\tlq^2/2)^2/2\tlq^2\tlt-2\tlpp^2/\tlt}  \times\cr
&& \biggl[ (\frac{\pi}{2})^2(\delta(v_2)-1)(C_{q2}^2-1)
+\bigl(\frac{1}{4v_2}\log \bigl(\frac{1+v_2}{1-v_2}\bigr)^2-Q_0(v_2)^2 \bigr)S_{q2}^2
+\pi \bigl( \frac{1}{2v_2} -Q_0(v_2) \bigr) C_{q2}S_{q2} \cr
&& +\ (\frac{\pi}{2})^2(\delta(v_3)-1)(C_{p2}^2-1)
+\bigl(\frac{1}{4v_3}\log \bigl(\frac{1+v_3}{1-v_3}\bigr)^2-Q_0(v_3)^2 \bigr)S_{p2}^2
+\pi \bigl( \frac{1}{2v_3} -Q_0(v_3) \bigr) C_{p2}S_{p2} 
\biggr]. \cr
&&
\eea
\end{widetext} 
Note again the 1st, 3rd, 4th, and 6th terms are nonzero only for $|v_2|$ or $ |v_3| \le 1$. Principal-vaule integrals are assumed where it is appropriate.
\item Spin parallel case
\begin{widetext}
\bea\label{bcs_uu}
\Delta  \tilde{S}^b_{\uparrow\uparrow,2}(q,\omega) &&=\ \int d\tlpp  d\tlqq  \frac{3\tlt}{4\tlq^3} (\frac{2}{\pi})^2 e^{-(\tlo+2\tlpp^2-2\tlqq^2-\tlq^2/2)^2/2\tlq^2\tlt-2\tlpp^2/\tlt}  \times\cr
&& \biggl[ -(\frac{\pi}{2})^2 P_0(v_2)(C_{q2}^2-1)
+\bigl(\frac{1}{v_2^2-1}-Q_0(v_2)^2 \bigr)S_{q2}^2
-\pi P_0(v_2)Q_0(v_2) C_{q2}S_{q2} \cr
&& -(\frac{\pi}{2})^2 P_0(v_3)(C_{p2}^2-1)
+\bigl(\frac{1}{v_3^2-1}-Q_0(v_3)^2 \bigr)S_{p2}^2
-\pi P_0(v_3)Q_0(v_3) C_{p2}S_{p2}
\biggr].
\eea
\end{widetext}
Similarly, the 1st, 3rd, 4th, and 6th terms are nonzero only for $|v_2|$ or $ |v_3| \le 1$. 
\end{enumerate}

On the BEC side, the contribution from transition between bound state and scattering state is obtained as follows,
\begin{widetext}
\bea\label{bec1}
 \Delta  \tilde{S}^d_{\sigma\sigma^{\prime},2}(q,\omega)
&=& \int d\tlqq  \frac{3\tlt}{4\tlq^3}e^{-(\tlo -2/k_F^2a^2-2\tlqq^2-\tlq^2/2)^2/2\tlq^2\tlt+2/\tlt k_F^2a^2}  \biggl[    
\left( A(u_2)\ -\ Q_0(u_2)^2 \right) \cr
&& + \left( Q_0(u_2)^2 C_{q_2} +(\arctan[(\tlq/2-\tlqq)k_Fa]+\arctan[(\tlq/2+\tlqq)k_Fa]) S_{q_2} \right)^2\biggr] \cr
&& + \int d\tlpp  \frac{3\tlt}{4\tlq^3}  e^{-(\tlo+2\tlpp^2+2/{k_F^2a^2}-\tlq^2/2)^2/2\tlq^2\tlt-2\tlpp^2/\tlt}  \biggl[    
\left( A(u_3)\ -\ Q_0(u_3)^2 \right) \cr
&& + \left( Q_0(u_3)^2 C_{p_2} +(\arctan[(\tlq/2-\tlpp)k_Fa]+\arctan[(\tlq/2+\tlpp)k_Fa]) S_{p_2} \right)^2
\biggr], 
\eea 
\end{widetext}
where
$u_2= \left[1/k_F^2a^2+\tlq^2/4+\tlqq^2\right] /\tlq\tlqq$,
$u_3= \left[1/k_F^2a^2+\tlq^2/4+\tlpp^2\right] / \tlq\tlpp$, and
$A(x)= 1/(x^2-1),({\rm for\ \uparrow\uparrow})$, 
or $Q_0(1/x)/x, ({\rm for\ \uparrow\downarrow}).
$

\section{Virial expansion of density to second order}
\label{app:app3}

On BCS side, the following equation for number density,
\bea
\frac{n\lambda^3}{2} &&=\ \frac{(4\pi/\tlt)^{3/2}}{6\pi^2}\
=\ \frac{2}{\sqrt{\pi}}\int_0^\infty dt \frac{\sqrt{t}}{1+e^t/z}\cr &&+\ \sqrt{2}z^2 e^{2/\tlt (k_Fa)^2} \mathrm{Erfc}[-\sqrt{\frac{2}{\tlt}}\frac{1}{k_Fa}],
\eea is used to solve fugacity $z$.

On BEC side, the equation for number density is changed accordingly due to the additional contribution from bound state, 
\bea
\frac{n\lambda^3}{2} &&=\ \frac{(4\pi/\tlt)^{3/2}}{6\pi^2}\
=\ \frac{2}{\sqrt{\pi}}\int_0^\infty dt \frac{\sqrt{t}}{1+e^t/z}\cr && +\ \sqrt{2}z^2 e^{2/\tlt (k_Fa)^2}\left( 2- \mathrm{Erfc}[\sqrt{\frac{2}{\tlt}}\frac{1}{k_Fa}] \right ).
\eea




\begin{thebibliography}{99}
\bibitem{foot1}
To be precise, one has to multiply a factor of $\frac 1 2$ when $v_2(v_3)=1$ in front of $P_l(v_2)$. It has negligible influence on the integral.
\end{thebibliography}
\end{document}